\documentstyle [12pt]{article}
\begin{document}
\baselineskip 20.0 pt
\par
\mbox{}
\vskip -1.25in\par
\mbox{}
 \begin{flushright}
\makebox[1.5in][1]{ITP-SB-92-65}\\
\makebox[1.5in][1]{UU-HEP-92/22}\\
\makebox[1.5in][1]{December 1992}
\end{flushright}
\vskip 0.25in

\begin{center}
{\large $D>2$ TOPOLOGICAL STRING}\\
\vspace{40 pt}
{Feng Yu}\\
\vspace{20 pt}
{{\it Institute for Theoretical Physics, State University of New York}}\\
{{\it Stony Brook, New York 11794, U.S.A.}}\\
{{\it and}}\\
{{\it Department of Physics, University of Utah}}\\
{{\it Salt Lake City, Utah 84112, U.S.A.}}\\
\vspace{40 pt}
{{\large Abstract}}\\
\end{center}
\vspace{10 pt}

A $D>2$ topological string is presented by coupling the $2d$ topological
gravity with the twisted version of the $N=2$ superconformal matter with
$c=3k/(k-2)$. The latter is shown to admit $k+1$ chiral primary fields from
the $SL(2,R)_{k}/U(1)$ unitary irreducible representations. The analysis of
topological contact interactions along with the consistency requirement
lead to recursion relations of correlation functions, that are convertable
to the Virasoro constraints on the perturbed partition function. It is further
expected to satisfy the nonlinear $\hat{W}^{(k-2)}_{\infty}$ constraints
associated with the graded algebra $SL(k,2)$, and thus the model is completely
solvable at arbitrary genus of the surface.

\newpage

{\it Introduction} --- There have been encouraging new developments of
string theory in last three years, led by the success of the matrix model
approach[1] to the $2d$ quantum gravity, or more generally, strings with
space-time dimension $D\leq 2$. The most distinguished feature is perhaps the
nonperturbative solvability of all correlation functions upto all orders of
Riemann surface genus, which is however best interpreted field-theoretically
in the equivalent topological model approach[2]. Whereas these $D\leq 2$
models present rich and beautiful structures both physically and
mathematically, in the context of string theory, they serve as toy models.
It is of essential interest to search for also exactly solvable string models
beyond the $D=2$ ``barrier'', where the topological phase is expected to
emerge as a transition from the ordinary phase. We study the former in this
Letter.

In $D<2$ case, the relevant $2d$ topological field theory[2,3,4] -- $2d$
topological gravity coupled to the twisted[5] $N=2$ superconformal minimal
matter with central charge $c=3k/(k+2)<3$ was shown[6] to be solved by
recursion relations of physical correlators, which by consistency coincide
with the extended conformal $W_{k+2}$ constraints on the perturbed partition
function. Counting in the Liouville modes
of gravity sector one space-time dimension, the coupled model is usually
interpreted as the $D=1+c/3$ topological string[7]. In the same sense,
topological gravity coupled to suitable $c>3$ conformal matter would lead to
the desired $D>2$ topological string.

There indeed exists a series of such $N=2$ superconformal matter with
$c=3k/(k-2)$[8]. Naturally, its twisted version becomes the relevant
topological matter, which we show in this Letter has (finitely) $k+1$
nontrivial BRST-invariant local operators, corresponding precisely to as many
chiral primaries[9] in the untwisted model. Then for the underlying $D>2$
topological string, by analysing topological contact interactions with
techniques of Ref.[4,6], slightly different recursion relations are obtained
and converted to the Virasoro constraints. The consistency of multi-contact
interactions involving higher-spin currents and the noncompactness of the
matter sector suggests that the complete set of recursion relations,
which determine all correlaton functions, are implied by the nonlinear
$W$-constraints of the infinite-type[10,11], in particular the
$\hat{W}^{(k-2)}_{\infty}$ constraints associated with the graded Lie algebra
$SL(k,2)$[12].

{\it $c>3$ Topological Matter} --- Consider the $N=2$ superconformal algebra,
generated by the stress tensor $T(z)$, the $U(1)$ current $J(z)$ and two
spin-3/2 fermionic currents $G^{\pm}(z)$. For central charge $c>3$, the $N=2$
algebra can be realized via the generalized parafermions $\Psi^{\pm}(z)$
plus a free boson $\phi (z)$[13,8], e.g.,
\begin{eqnarray}
G^{\pm} = \sqrt{\frac{2c}{3}}\Psi^{\pm}e^{\pm i\sqrt{3/c}\phi}, ~~~
J = i\sqrt{\frac{c}{3}}\partial\phi;
\end{eqnarray}
so does the $SL(2,R)_{k}$ current algebra with
\begin{eqnarray}
J^{\pm} = \sqrt{k}\Psi^{\pm}e^{\pm i\sqrt{2/k}\phi}, ~~~
J^{3} = -\sqrt{\frac{k}{2}}\partial\phi,
\end{eqnarray}
where $c=3k/(k-2)$ and the level $k (>2)$ is restricted to integer in the
scope of this Letter. Hence the irreducible representations of the former
may be obtained from that of the latter. Notice the signatures of $J(z)$ in
(1) and $J^{3}(z)$ in (2) are opposite. So although the $SL(2,R)$ module is
not positive-definite because of its noncompact nature, the $N=2$ module
becomes unitary as required.

Recall there are two discrete ${\it D}^{\pm}$ and one continuous $C$ series
of irreducible $SL(2,R)$ representations. Indicate the lowest (highest)
eigenstate of $J^{3}_{0}$ mode in $D^{+}$ ($D^{-}$) as $|l\rangle$ with $l$
being positive (negative) integers or half-integers. From Ref.[8], the
unitarity region for the coset $SL(2,R)_{k}/U(1)$ such that the $N=2$
algebra by adding back a free boson with different signature is
\begin{eqnarray}
k\geq 2|l|\geq 0, ~~~k>2;
\end{eqnarray}
and in the $C$ series, simply $k>2$. Let the primary states of the $N=2$
representations, e.g., in the $D^{\pm}$ series, labeled by $|\phi_{l,n}\rangle$
$(n=0,1,\ldots)$. Note $|\phi_{l,n}\rangle$ are annihilated by positive fermion
modes $G^{\pm}_{n+1/2}$, and the dimensions from the zero mode of $T(z)$
are
\begin{eqnarray}
h = (n+\frac{1}{2})|q|-\frac{n(n+1)}{k-2}
\end{eqnarray}
with the $U(1)$ charges $q=-2(l\pm n)/(k-2)$[8]. The solution of the
(anti-)chiral primary states with $h=\pm q/2$
(or $G^{\pm}_{-1/2}|\phi_{l,n}\rangle =0$) turns out to be $n=0$. Thus,
there are finitely many either chiral
or anti-chiral primaries in $D^{-}$ or $D^{+}$ series bounded by (3).
Meanwhile there is no such state in the continuous series.

Upon twisting the $N=2$ algebra (in the NS sector): $T(z)\rightarrow
T(z)+(1/2)\partial J(z)$, $G^{+}_{-1/2}$ becomes a BRST charge $Q$ with spin 1
while $G^{-}(z)$ a spin-2 current $G(z)$, such that the stress tensor
$T(z)=-{\{Q,G(z)\}}$ is a BRST commutator, giving the topological nature
of the twisted model. The local physical operator $V_{\alpha}$
that are nontrivially BRST invariant, are precisely the chiral priamry
fields in the untwisted model therefore have ghost charges
\begin{eqnarray}
q_{\alpha} = \frac{\alpha}{k-2}, ~~~~\alpha = 0,1,\ldots, k.
\end{eqnarray}
It is easy to extend the field-theoretical realization of $c<3$ topological
minimal matter with a pair of free bosons $(\varphi,\bar{\varphi})$ and
free fermions $(\lambda,\bar{\lambda})$[6] to the current $c>3$ case, in
which the (zero-form) primary operators are of the form
\begin{eqnarray}
V_{\alpha} = e^{-\frac{\alpha}{\sqrt{k-2}}\varphi}.
\end{eqnarray}
The correlators of (6) and their one-form partners give rise to physical
amplitudes of the underlying topological matter.

{\it $D>2$ Topological String} --- In turn, consider the conformal invariant
description of $2d$ topological gravity[4]. Its field contents consist of the
Liouville pair $(\phi,\pi)$, their supersymmetric partner $(\psi,\chi)$ and
the ghost pairs $(b,c)$, $(\bar{b},\bar{c})$, $(\beta,\gamma)$,
$(\bar{\beta},\bar{\gamma})$ from gauge-fixing. The basic BRST-invariant
(zero-form) physical operators are of the form
\begin{eqnarray}
\sigma_{n} = \gamma_{0}^{n} = \frac{1}{2^{n}}(\partial\gamma +
\gamma\partial\phi + c\partial\psi - c.c.)^{n}, ~~~n\geq 0,
\end{eqnarray}
and have ghost dimension $2n$. An important feature[2,3,4] of $2d$ topological
gravity is that (7) must be associated with a puncture operator
\begin{eqnarray}
P = c\bar{c}\delta(\gamma)\delta(\bar{\gamma})
\end{eqnarray}
of ghost dimension $-2$. It creates two bosonic and two fermionic zero modes
to give the right volume form on the moduli space of the punctured Riemann
surfaces, so that there emerge non-vanishing correlation functions of local
operators (7) times (8), representing the intersection numbers of the
cycles on the moduli space[14,15].

The $D>2$ topological string naturally appears as the topological gravity
coupled with the above $c>3$ topological matter, of which a covariant
field-theoretical formalism can be obtained through direct extension of the
$c<3$ case[6]. Now the complete set of nontrivial local physical operators,
that are invariant under the total BRST operation of the coupled system,
read
\begin{eqnarray}
\sigma_{n,\alpha} = V_{\alpha}\cdot\sigma_{n}\cdot P
\end{eqnarray}
with $V_{\alpha}$ a covariantized version ($\varphi\rightarrow
\varphi-c\lambda$, etc.) of (6), whose ghost charge (both left- and
right-moving parts of the matter sector included) is $2n-2+2\alpha/(k-2)$.
The correlation functions of operators (9)
\begin{eqnarray}
\langle \sigma_{n_{1},\alpha_{1}}\sigma_{n_{2},\alpha_{2}}\cdots
\sigma_{n_{s},\alpha_{s}} \rangle_{g}
\end{eqnarray}
are defined as path integrals of them over the surface with $g$ handles and
$s$ punctures and over the (supersymmetric) moduli space.
The selection rule for non-vanishing values of (10) is just the ghost charge
conservation. Taking into account the background charge of gravity sector
associated with the supersymmetric moduli $-6(g-1)$ and that of the matter
sector from the $U(1)$ ghost current $2(g-1)c/3$, it reads
\begin{eqnarray}
\sum^{s}_{i=1}(n_{i}-1+\frac{\alpha_{i}}{k-2}) = 2(g-1)\frac{k-3}{k-2}.
\end{eqnarray}
In the meantime, the background charge of Liouville fields is cancelled by
the BRST-invariant insertion of $\delta$-function-surface-curvature creating
operators $\prod_{i}e^{q_{i}(\prod -c\partial\chi -\bar{c}\bar{\partial}\chi)}$
with $\sum_{i}q_{i}=2g-2$ to the path integrals.

The amplitudes (10) are independent of the positions of operators, by the
topological nature of the model. The integrals over these puncture positions
(the moduli created by the punctures) do not contribute except in the region
where punctures are very close or at least one puncture approaches
a node on the surface (caused by the satble compactification of the moduli
space). From field-theoretical point of view, they constitute the so-called
contact interactions. The analysis of them leads to a set of recursion
relations for (10) in terms of genus $g$ and the number of involved operators
$s$. Let us simply present these relations with brief explanation. Detailed
derivation is parallel to that in Ref.[4,6] where the techniques were
developed.

First, the puncture operator $P$ remains to behave as
$\delta$-fuction in the neighbourhood $\epsilon$ of another operator
$\sigma_{n,\alpha}$: $\int_{\epsilon}P|\sigma_{n,\alpha}\rangle =
|\sigma_{n-1,\alpha}\rangle$, such that
\begin{eqnarray}
\int_{\epsilon}\sigma_{m+1,0}|\sigma_{n,\alpha}\rangle =
|\sigma_{m+n,\alpha}\rangle.
\end{eqnarray}
Let the insertion of each curvature singularity $e^{q_{n,\alpha}(\prod
-c\partial\chi -\bar{c}\bar{\partial}\chi)}$ with $q_{n,\alpha}=
((k-2)(n-1)+\alpha)/(k-3)$ be right at the position of $\sigma_{n,\alpha}$,
the coefficient on the right side is then shifted to $\sigma_{n,\alpha}+1$.
Note by selection rule (11) alone, there could also exist a state
$|\sigma_{m+n+1,\alpha-k+2}\rangle$ on the right side when $\alpha\geq k-2$.
However, because the interaction from $\sigma_{m+1,0}$ is purely
gravitational, no such state emerges. Eq.(12) gives rise to the contact terms
$\langle \sigma_{m+1,0}\prod_{i}\sigma_{n_{i},\alpha_{i}}
\rangle_{\epsilon_{j}}$
$\sim$ $\langle \sigma_{m+n_{j},\alpha_{j}}\prod_{i\neq j}\sigma_{n_{i},
\alpha_{i}}\rangle$, each of that represents the formation of a node
separating two approaching points located on a sphere from the rest of the
surface. The other type of contact interaction, occuring when a puncture is
in the neighbourhood $\delta$ of a node, yields factorization terms
$\langle \sigma_{m+1,0}\prod_{i\in S}\sigma_{n_{i},\alpha_{i}}
\rangle_{\delta,g}$ $\sim$
$\langle \sigma_{m_{1},\alpha}\sigma_{m_{2},\beta}\prod_{i\in S}
\sigma_{n_{i},\alpha_{i}}\rangle_{g-1}$ or $\langle \sigma_{m_{1},\alpha}
\prod_{i\in X}\sigma_{n_{i},\alpha_{i}}\rangle_{g'}
\langle \sigma_{m_{2},\beta}
\prod_{j\in Y}\sigma_{n_{j},\alpha_{j}}\rangle_{g-g'}$ $(S=XUY)$. Both indicate
the formation of a second node, but the former breaks a handle whereas the
latter divides the surface into two parts. Summing up these ingredients
completes the integral over $\sigma_{m+1,0}$'s position, giving rise to the
desired recursion relations:
\begin{eqnarray}
\langle \sigma_{m+1,0}\prod_{i\in S}\sigma_{n_{i},\alpha_{i}}\rangle_{g}
&=& \frac{k-2}{k-3}\sum_{i\in S}(n_{i}+\frac{\alpha_{i}-1}{k-2})
\langle \sigma_{m+n_{i},\alpha_{i}}\prod_{j\neq i}\sigma_{n_{j},\alpha_{j}}
\rangle_{g} \nonumber\\
& & +\sum_{n=1}^{m}\sum_{\alpha =0}^{k}(b_{n,\alpha}^{m}
\langle \sigma_{n-1,\alpha}\sigma_{m-n,k-\alpha}\prod_{i\in S}
\sigma_{n_{i},\alpha_{i}}\rangle_{g-1} \\
& & +\sum_{S=XUY,g'}d_{n,\alpha}^{m}
\langle \sigma_{n-1,\alpha}\prod_{i\in X}\sigma_{n_{i},\alpha_{i}}\rangle_{g'}
\langle \sigma_{m-n,k-\alpha}\prod_{j\in Y}\sigma_{n_{j},\alpha_{j}}
\rangle_{g-g'}). \nonumber
\end{eqnarray}
{}From the selection rule (11), more factorization terms, such as
$\langle \sigma_{n,\alpha}\sigma_{m-n,2-\alpha}\prod_{i}
\sigma_{n_{i},\alpha_{i}}\rangle_{g-1}$, are permitted; but they are not
compatible with the contact terms, thus eliminated. Eq.(13) solves many
correlation functions recursively upto the metric
\begin{eqnarray}
\eta_{\alpha\beta} \equiv \langle \sigma_{0,0}\sigma_{0,\alpha}
\sigma_{0,\beta} \rangle_{0} = \delta_{\alpha+\beta,k}A_{\alpha}
\end{eqnarray}
and the one point function on the torus
\begin{eqnarray}
\langle \sigma_{1,0} \rangle_{1} = C.
\end{eqnarray}
The consistency of all equations in (13) and with (14)-(15), especially the
requirement that positions of all operators in physical correlators are
symmetric, determines all coefficients upto the normalization of $k+1$
primaries $\sigma_{0,\alpha}$:
\begin{eqnarray}
b_{n,\alpha}^{m}=d_{n,\alpha}^{m}=\frac{k-2}{2(k-3)},~ A_{\alpha}=
\frac{(\alpha-1)(k-\alpha-1)}{(k-2)(k-3)},~
C=\frac{(k^{2}-1)(k-6)}{24(k-2)(k-3)}.
\end{eqnarray}
Nevertheless, (13)-(16) are only partial solutions corresponding to two-point
contact interactions initiated by pure gravity operator $\sigma_{m,0}$ and
therefore involving spin-2 currents. The complete set of recursion relations
correspond to multi-contact interactions of general $\sigma_{m,\alpha}$,
which behave as higher-spin currents. Let us discuss them from the following
constraint point of view.

{\it Virasoro and $W$-Constraints} --- Consider the model's partition function
\begin{eqnarray}
Z(t) = \langle e^{\sum_{n,\alpha}t_{n,\alpha}\sigma_{n,\alpha}} \rangle
\equiv \sum_{g}\lambda^{2g-2}\langle e^{\sum_{n,\alpha}t_{n,\alpha}
\sigma_{n,\alpha}} \rangle_{g}
\end{eqnarray}
perturbed by the couplings $t_{n,\alpha}$. It generates all unperturbed
correlation functions
\begin{eqnarray}
\langle \prod^{s}_{i=1}\sigma_{n_{i},\alpha_{i}} \rangle =
\frac{\partial^{s}}{\partial t_{n_{1},\alpha_{1}}\partial t_{n_{2},\alpha_{2}}
\cdots \partial t_{n_{s},\alpha_{s}}} \log Z(t).
\end{eqnarray}
The recursion relations (13)-(16) can be converted into the following
Virasoro constraints
\begin{eqnarray}
L_{m}Z(t) = 0, ~~~~ m\geq -1
\end{eqnarray}
with
\begin{eqnarray}
L_{-1} &=& \sum^{\infty}_{n=1}\sum^{k}_{\alpha =0}(n+\frac{\alpha -1}{k-2})
t_{n,\alpha}\frac{\partial}{\partial t_{n-1,\alpha}} + \frac{1}{2}
\lambda^{-2}\sum^{k}_{\alpha =0}\frac{(\alpha -1)(k-\alpha -1)}{(k-2)^{2}}
t_{0,\alpha}t_{0,k-\alpha}, \nonumber\\
L_{0} &=& \sum^{\infty}_{n=0}\sum^{k}_{\alpha =0}(n+\frac{\alpha -1}{k-2})
t_{n,\alpha}\frac{\partial}{\partial t_{n,\alpha}} +
\frac{(k^{2}-1)(k-6)}{24(k-2)^{2}}, \\
L_{m} &=& \sum^{\infty}_{n=0}\sum^{k}_{\alpha =0}(n+\frac{\alpha -1}{k-2})
t_{n,\alpha}\frac{\partial}{\partial t_{n+m,\alpha}} + \frac{1}{2}
\lambda^{2}\sum^{m}_{n=1}\sum^{k}_{\alpha =0}\frac{\partial^{2}}{\partial
t_{n-1,\alpha}\partial t_{m-n,k-\alpha}}, \nonumber
\end{eqnarray}
expanded around the critical point $t_{n,\alpha}=-\delta_{n,1}
\delta_{\alpha,0}$. Note by these constraints the solutions can globally
flow into other critical points corresponding to more general (topological)
models.

Assuming (13)-(16) or (19)-(20) has a unique extension to the whole solution,
the complete set of recursion relations are then equivalent to the
(sufficiently large) extended $W$-constraints on $Z(t)$. To search for the
latter, recall in the $D<2$ case[6], the fact that the Virasoro constraints
are the negative modes of the stress tensor realized with $k+1$ free bosons
twisted by multiples of $1/(k+2)$ identifies the latter as that of level-1
$SL(k+2)$ current algebra. The relevant $W$-constraints then turn out to be
the $SL(k+2)$ associated $W_{k+2}$ algebra. For the present $D>2$ case, (20)
are nothing but the negative modes of the $c=k+1$ stress tensor
\begin{eqnarray}
T(z) = \sum^{k}_{\alpha =0}\tilde{j}_{\alpha}(z)\tilde{j}_{k-\alpha}(z)
+ \frac{(k^{2}-1)(k-6)}{24(k-2)^{2}z^{2}}
\end{eqnarray}
in which $\tilde{j}_{\alpha}=\partial \tilde{\phi}_{\alpha}=\sum_{n}
\tilde{j}_{n,\alpha}z^{-n-1-(\alpha -1)/(k-2)}$ are the free currents twisted
by $(\alpha -1)/(k-2)$, with modes identified with $\tilde{j}_{n,\alpha}=
\lambda\partial/\partial t_{n,\alpha}$, $\tilde{j}_{-(n+1),k-\alpha}=
\lambda^{-1}(n+\frac{\alpha -1}{k-2})t_{n,\alpha}$. Note it is straightforward
to show that the weight vectors of the fundamental representation of graded
Lie algebra $SL(p,q)$ have inner product
\begin{eqnarray}
\vec{h}_{i}\cdot\vec{h}_{j} = \left\{\begin{array}{ll}
\delta_{ij}-1/(p-q), & i,j=1,\ldots,p\\
1/(p-q), & i=1,\ldots,p, ~j=p+1,\ldots,p+q\\
-\delta_{ij}-1/(p-q), & i,j=p+1,\ldots,p+q,
\end{array} \right.
\end{eqnarray}
matching the $1/(k-2)$-twisting with $p=k,q=2$. Thus the level-1 $SL(k,2)$
current algebra appears to generate the relevant $c=k+1$ $W$-algebra with
(21) the Virasoro current, via standard reduction. The negative modes of
this $W$-algebra's currents consistently serve as the desired $W$-constraints
\begin{eqnarray}
W_{r,m}Z(t) =0, ~~~~ r\geq 2, ~m\geq -r+1,
\end{eqnarray}
which are expected to exactly solve the $D>2$ topological string.

{\it $\hat{W}^{(p-q)}_{\infty}$ and $SL(p,q)$} --- Let us further identify the
$W$-algebra associated with $SL(k,2)$. We proceed more generally. Recall it
was shown[10] that there exists a nonlinear algebra $\hat{W}_{\infty}$,
or even $\hat{W}_{1+\infty}$ by incorporating a spin-1 current, whose
structure is most concisely exibited as the second Hamiltonian structure
of the integrable KP hierarchy with the spin-($r+2$) generators defined as
the coefficient function $u_{r}(z)$ in the pseudo-differential Lax
operator[16]:
\begin{eqnarray}
L = D + \sum^{\infty}_{r=-1}u_{r}D^{-r-1}, ~~~~D\equiv \partial/\partial z.
\end{eqnarray}
Moreover, similar manipulation in terms of the powers of (24)
\begin{eqnarray}
L^{N} = D^{N} + \sum^{\infty}_{r=-N}v_{r}D^{-r-1}
\end{eqnarray}
results in non-equivalent Hamiltonian structures $\hat{W}^{(N)}_{1+\infty}$
with $v_{r}(z)$ the spin-($r+N+1$) generators[11]. The
$\hat{W}^{(N)}_{1+\infty}$ ($\hat{W}^{(1)}_{1+\infty}\equiv\hat{W}_{1+\infty}$)
algebra is proved[12] to admit an infinite number of free boson
representations that $v_{r}(z)$ are realized through
\begin{eqnarray}
L^{N} = \prod^{p}_{i=1}(D+j_{i})\prod^{p+q}_{l=p+1}(D-j_{l})^{-1}, ~~~~p-q=N,
\end{eqnarray}
as functions of free currents $j_{i}(z)=\partial\phi_{i}(z)$ and
$j_{l}(z)=\partial\phi_{l}(z)$ of the opposite signature and their derivatives,
generalizing the Miura transformation[17] for $W_{N}$ and two boson
realization[18] for $\hat{W}_{\infty}$. By imposing the (second-class)
constraint $v_{-N}=\sum_{i}j_{i}+\sum_{l}j_{l}=0$, $\hat{W}^{(N)}_{1+\infty}$
is reduced to $\hat{W}^{(N)}_{\infty}$. Correspondingly, (26) is modified to
\begin{eqnarray}
L^{N} = \prod^{p}_{i=1}(D+\vec{h}_{i}\cdot\vec{j})\prod^{p+q}_{l=p+1}
(D-\vec{h}_{l}\cdot\vec{j})^{-1}
\end{eqnarray}
where $\vec{j}= (j_{1},\cdots,j_{p},j_{p+1},\cdots,j_{p+q})$ and $\vec{h}_{i}$
are precisely the fundamental weight vectors of $SL(p,q)$ satisfying (22).
Hence (27) is the Lax operator associated with $SL(p,q)$, via bosonic
realization. Furthermore, the quantization of $\hat{W}^{(p-q)}_{\infty}$
may be carried out through, for instance, the Drinfeld-Sokolov reduction[19]
from the level-1 graded $SL(p,q)$ current algebra. In particular,
$\hat{W}^{(k-2)}_{\infty}$ fits the role of the relevant $W$-algebra from
$SL(k,2)$.

{\it Remark} --- Finally, the exact solvability of the underlying $D>2$
topological string suggests a close connection to complete integrable
differential systems. Of prime consideration is the KP hierarchy. We
speculate that the perturbed partition function (17) is a KP $\tau$-function
on its solution space[16], extending the wonderful outcomes at
$D<2$[2,3,7,14,15] but within the Witten's general conjecture[2,14]. This
$\tau$-function is further expected to be written in terms of the
Kontsevich-type matrix integrals[15], from which the fashion of
$\hat{W}^{(k-2)}_{\infty}$ constraints would be more transparent.

{\it Acknowledgements}:~ The auther thanks J. Lykken, M. Rocek, W. Siegel
and Y.-S. Wu for discussions, and particularly C.N. Yang for providing
a stimulating atmosphere and support. This work is supported in part by NSF
grant PHY-9008452.

\vspace{10 pt}
\begin{center}
{\large References}
\end{center}
\begin{itemize}

\item[1.] D. Gross and A. Migdal, Phys. Rev. Lett. 64 (1990) 127; M. Douglas
and S. Shenker, Nucl. Phys. B335 (1990) 635; E. Brezin and V. Kazakov, Phys.
Lett. B236 (1990) 144.
\item[2.] E. Witten, Nucl. Phys. B340 (1990) 281.
\item[3.] R. Dijkgraaf and E. Witten, Nucl. Phys. B342 (1990) 486.
\item[4.] E. Verlinde and H. Verlinde, Nucl. Phys. B348 (1991) 457.
\item[5.] E. Witten, Commun. Math. Phys. 117 (1988) 353; T. Eguchi and
S.-K. Yang, Mod. Phys. Lett. A5 (1990) 1693.
\item[6.] K. Li, Nucl. Phys. B354 (1991) 711; 725.
\item[7.] R. Dijkgraaf, E. Verlinde and H. Verlinde, Nucl. Phys. B352 (1991)
59.
\item[8.] L. Dixon, M. Peskin and J. Lykken, Nucl. Phys. B325 (1989) 329.
\item[9.] W. Lerche, C. Vafa and N.P. Warner, Nucl. Phys. B324 (1989) 427.
\item[10.] F. Yu and Y.-S. Wu, Nucl. Phys. B373 (1992) 713; Utah preprint
UU-HEP-92/12.
\item[11.] J. M. Figueroa-O'Farrill, J. Mas and E. Ramos, preprint
BONN-HE-92/20; 92/$\alpha$.
\item[12.] F. Yu, to appear.
\item[13.] J. Lykken, Nucl. Phys. B313 (1989) 473.
\item[14.] E. Witten, Surveys In Diff. Geom. 1 (1991) 243.
\item[15.] M. Kontsevich, Funk. Anal. i Pril. 25 (1991) 50.
\item[16.] M. Sato, RIMS Kokyuroku 439 (1981) 30; E. Date, M. Jimbo, M.
Kashiwara and T. Miwa, in Proc. of RIMS Symposium on Nonlinear Integrable
Systems, eds. M. Jimbo and T. Miwa, (World Scientific, Singapore, 1983);
G. Segal and G. Wilson, Publ. IHES 61 (1985) 1.
\item[17.] V. Drinfel'd and V. Sokolov, Sov. Probl. Math. 24 (1984) 81.
\item[18.] F. Yu and Y.-S. Wu, Phys. Rev. Lett. 68 (1992) 2996.
\item[19.] V. Drinfel'd and V. Sokolov, Journ. Sov. Math. 30 (1985) 1975.

\end{itemize}

\end{document}